\documentclass{PoS}

\def\f#1   {Fig.~\ref{#1}}
\def\s#1   {Sect.~\ref{#1}}
\def\tab#1   {Table~\ref{#1}}
\def\eq#1   {Eq.~\ref{#1}}
\def\t#1   {Table~\ref{#1}}
\def\lum   {$\mathrm{L}_\mathrm{1.4GHz}$}
\def\comm#1   {{\tt (COMMENT: #1) }}

\def\wh                {W~Hz$^{-1}$}

\def\smo	{Smol\v{c}i\'{c}}

\title{Radio continuum surveys and galaxy evolution: The AGN view}

\ShortTitle{Radio AGN and their cosmic evolution}

\author{\speaker{Vernesa Smol\v{c}i\'{c}}\thanks{Funded by the European Union's Seventh Framework programme 
under grant agreement 337595.}\\
        Department of Physics, University of Zagreb, Bijeni\v{c}ka cesta 32, HR-10000 Zagreb, Croatia\\
        E-mail: \email{vs@phy.hr}}

\abstract{
Understanding how galaxies form in the early universe and their subsequent evolution through cosmic time is a major goal of modern astrophysics. Panchromatic look-back sky surveys significantly advanced the field in the past decades, and we are now entering an even more fruitful period - a 'golden age' of radio astronomy - with upgraded, and new facilities delivering an order of magnitude increase in sensitivity. An overview of recent developments in radio continuum sky surveys, focusing on the physical properties and cosmic evolution of radio AGN since $z\sim5$ is presented here. 
}

\FullConference{EXTRA-RADSUR2015 (*)\\
		20--23 October 2015\\
		Bologna, Italy

                \bigskip
                \hrule
                \bigskip

                \textnormal{(*) This conference has been organized
                  with the support of the Ministry of Foreign Affairs
                  and International Cooperation, Directorate General
                  for the Country Promotion (Bilateral Grant Agreement
                  ZA14GR02 - Mapping the Universe on the Pathway to
                  SKA)}
}

\begin{document}

\section{Introduction}

Understanding how galaxies form in the early universe and their subsequent evolution through cosmic time is a major goal of modern astrophysics. 
This field has been significantly advanced in the past decade by panchromatic look-back sky surveys. The results of these, in synergy with those based on theoretical and numerical studies led 
 towards a picture of galaxy evolution in which blue star forming galaxies with spiral morphology evolve over time towards quiescent red galaxies with spheroidal morphologies and the highest
stellar masses (e.g., Faber et al. 2007). A galaxy evolves through interspersed episodes of intense
mass accretion onto the stellar body, as well as the central super-massive black hole (SMBH),
creating a powerful active galactic nucleus (AGN; Sanders \& Mirabel 1996).   In this context, understanding radio luminous AGN and their evolution through cosmic time is highly important. The radio outflows of such AGN are by now regularly taken in cosmological models as key to limit the growth of stellar mass in galaxies (so-called radio-mode feedback; Granato et al. 2004; Bower et al. 2006; Croton et al. 2006).  They are assumed to heat the gas within the galaxy and thereby prevent further star formation (SF), and thus further growth of the galaxy's stellar body. This allows the models to reproduce the observational results at the high-mass end, solving a long-standing problem in cosmological models of overpredicting the volume densities of the most massive galaxies in the Universe  (e.g., Somerville \& Primack 1999). However, to-date, it is still unclear whether such a process indeed occurs in nature with the assumed impact, further motivating studies of the properties of radio AGN, their cosmic evolution, impact on  environment and  role in galaxy formation and evolution (e.g., Hickox et al.\ 2009; \smo \ et al.\ 2015; Godfrey \& Shabala 2016). 
 
Numerous past studies have revealed that radio AGN predominantly appear as two types. However,  various different classification schemes of radio AGN, based on their observational properties have been suggested and applied onto radio AGN in the literature. These are based on e.g.\ radio morphology and luminosity -- FR~I vs.\ FR~II (Fanaroff \& Riley 1974; Ledlow \& Owen 1996), radio loudness -- radio-loud vs.\ radio-quiet AGN (e.g., Ivezi\'{c} et al.\ 2002, Balokovi\'{c} et al.\ 2011), radio spectrum -- steep vs.\ flat-spectrum sources (e.g., Dunlop \& Peacock 1990), or optical emission line properties -- low-excitation vs. high-excitation radio AGN (Hine \& Longair 1979; Laing et al. 1994) or, alternatively Seyferts, and quasars vs.\ LINERs and absorption line galaxies (e.g., \smo \ 2009). In the last decade the community has converged towards the last classification scheme due to a rise of evidence  supporting the idea that low- and high-excitation radio AGN (hereafter LERAGN and HERAGN, respectively) reflect fundamental physical differences, i.e., different modes of SMBH accretion, as well as differences in physical properties of their host galaxies. Hence, hereafter, the classification, properties and cosmic evolution of radio AGN will be analyzed and discussed in the context of LERAGN and HERAGN.

Significant advancements have been made in understanding the properties of radio AGN in the local ($z<0.3$) Universe thanks to
radio continuum surveys combined with surveys at other wavelengths. With the onset of upgraded facilities, these studies can be expanded to high redshift, allowing for the first determinations of the cosmic evolution of LERAGN- and HERAGN-dominated samples, and setting the path for future projects planned with the upgraded and new facilities, such as the Square Kilometre Array (SKA) and its precursors. This is summarized in \f{fig:surveys} \ where past, current, and future centimeter radio continuum surveys are put in context in terms of sensitivity vs.\ area. While previous surveys ranged from shallow surveys of thousands of square degrees on the sky (e.g., FIRST, NVSS, SUMMS; Becker et al.\ 1995; Condon et al.\ 1998; Bock et al.\ 1999) to the deepest surveys, but of single pointings only (e.g., Deep-SWIRE, Owen \& Morrison 2008),  for the same areas next generation surveys will push the limits  in sensitivity by multiple orders of magnitude  (e.g., SKA1 Deep and All Sky surveys; Prandoni et al.\ 2015). The trend of increasing sensitivity with new, intermediate- and wide-area surveys is also shown in \f{fig:surveys} \ in the context of radio luminosity vs.\ redshift. Various galaxy populations that can be identified in these surveys in combination with multi-wavelength data are also shown. This illustrates that these surveys will be able to generate statistically-significant samples of the faintest radio sources, such as faint AGN and SF galaxies, out to the highest redshifts. For example, an AGN with \lum~$\approx10^{24}$~\wh \ is detectable in the large-area shallow surveys (SDSS/NVSS/SUMSS) out to $z\lesssim0.5$. This limit was pushed by e.g.\ the 2 square degree VLA-COSMOS 1.4GHz Large Project (Schinnerer et al.\ 2007) to a limit of $z\lesssim2$, while the current VLA-COSMOS 3GHz Large Project (\smo \ et al., in prep) extended this to $z\lesssim3.5$. 

In summary, in synergy with good multi-wavelength coverage current radio surveys are probing statistically significant samples of fainter and higher-redshift radio sources than possible with previous surveys linking  past and future surveys in terms of sensitivity vs.\ area. This will allow for new insights into the properties and cosmic evolution of LERAGN and HERAGN out to redshifts of $\sim5$.
Our current understanding of these is summarized below. In \s{sec:class} \ various classifications of radio AGN are presented. In \s{sec:props} \ and \s{sec:evolv}  \ the properties of low- and high-redshift LE- and HERAGN, and their cosmic evolution are described, respectively. A summary and outlook is presented in \s{sec:summ} .

\begin{figure}
\begin{center}
\includegraphics[bb=  70 0 602 806, scale=0.35]{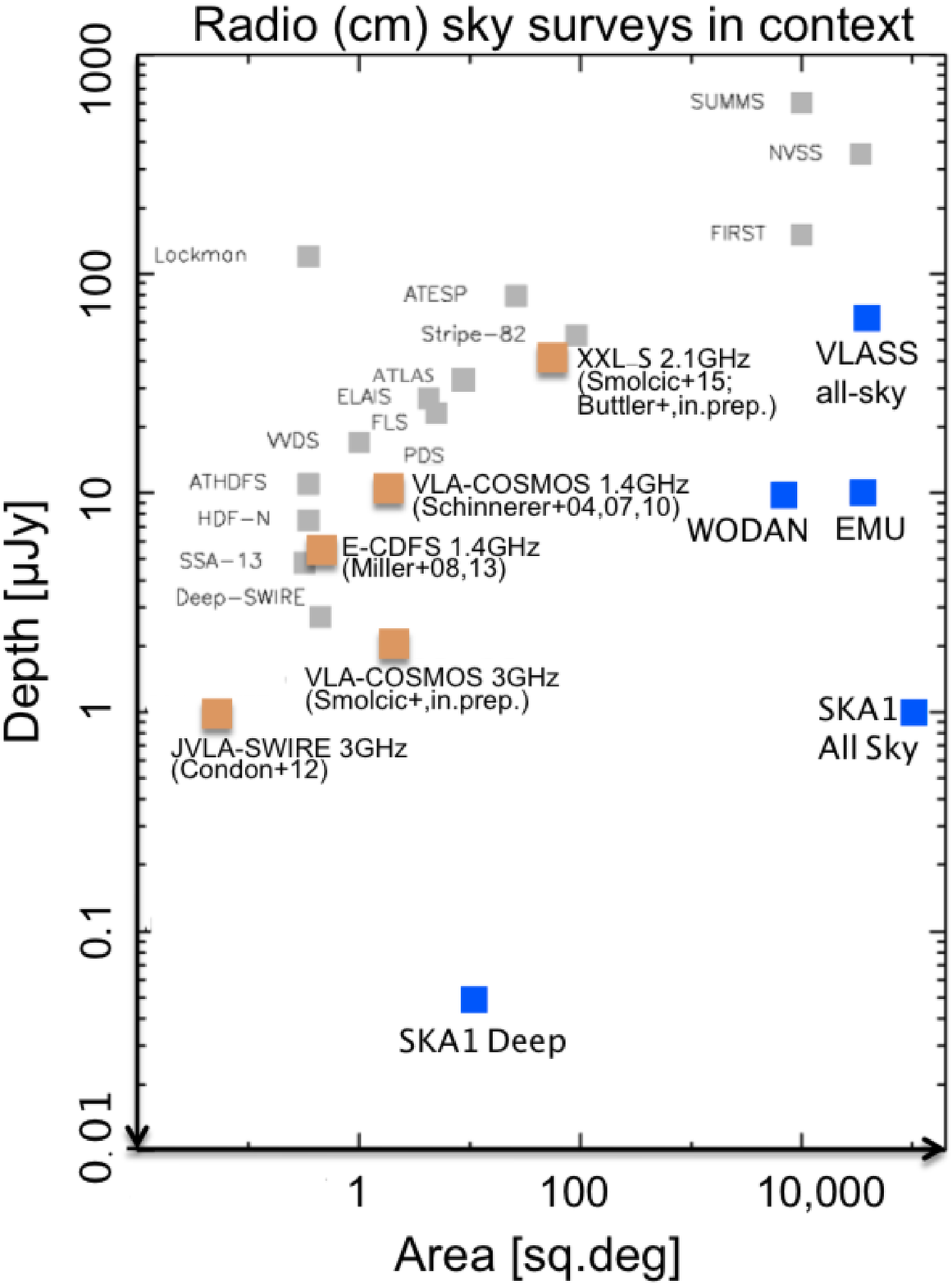}
\includegraphics[bb=  0 0 368 250, scale=0.65]{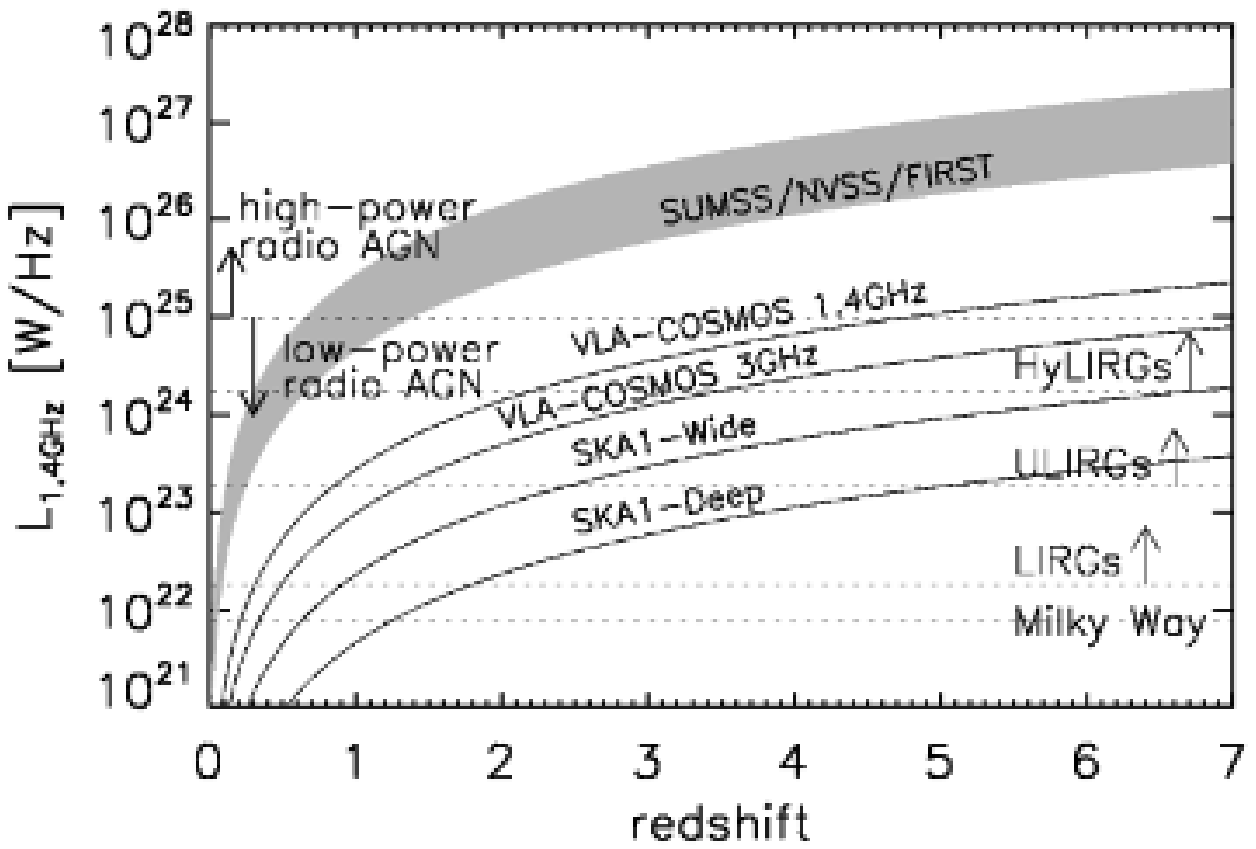}
\caption{Left panel: Past (gray symbols), current (orange symbols) and planned (blue symbols) radio continuum surveys in context of $1\sigma$ sensitivity vs.\ area. Right panel: Radio luminosity as a function of redshift with various survey limits and galaxy populations indicated. }
\label{fig:surveys}
\end{center}
\end{figure}

\section{Classification of radio AGN}
\label{sec:class}

The classification of radio AGN into HE- and LERAGN has been proposed by Hine \& Longair (1979). It is based on the
existence of high-excitation (HE) emission lines in the optical spectra of their host galaxies (Hine \& Longair 1979; Laing et al.\ 1994). In this scheme, objects without high-excitation emission lines are referred to as low-excitation radio
galaxies. In the context of one of the most commonly found separation of radio AGN in the literature, that into quasars, Seyfert, LINER, and absorption line galaxies (see \f{fig:bpt} ; Baldwin et al.\ 1981; Kauffmann et al. 2003; Kewley et al.\ 2001, 2006),  quasars and Seyferts would belong to the HERAGN category, while LINERs and absorption line galaxies would occupy the LERAGN class. Compared to the morphological classification of radio AGN into FR~I and II radio galaxies (Fanaroff \& Riley 1974; Ledlow \& Owen 1996), almost all FR I radio galaxies are LERAGN, while optical
hosts of the most powerful radio sources, i.e., FR IIs, usually have strong emission lines, and would, thus, be classified as HERAGN.
However, the correspondence between the FR class and the presence of emission lines is not one-to-one as many FR II galaxies have been found to be LERAGN (e.g., Evans et al. 2006). Comparing the HE-/LERAGN classification scheme with radio AGN separated into radio-loud and radio-quiet AGN is  more complicated given the various definitions of radio loudness in the literature (see Balokovi\'{c} et al.\ 2011 for a summary and parametrization of the most common radio-loudness definitions). If radio loudness is defined solely based on radio luminosity then radio-loud (radio-quiet) AGN samples would be dominated by HERAGN (LERAGN). However, for a definition of radio loudness taken as the ratio of radio to optical or (near-, mid-, far-, or monochromatic-) IR luminosity,  the reverse situation would occur.  Radio-loud (radio-quiet) AGN would be roughly consistent with the properties of LERAGN (HERAGN; see e.g., Bonzini et al.\ 2013 and Padovani et al.\ 2015 for more details). 

Using optical spectroscopic diagnostic tools to classify radio AGN into samples dominated by LERAGN or HERAGN is possible out to $z\sim1$, beyond which one has to rely on multi-wavelength proxies. This is due to a combination of i) the relevant  emission lines dropping-out of the wavelength range for a typically observed optical spectrum, and ii) the lack of complete spectroscopic coverage of deep photometric surveys (such as e.g. COSMOS, Scoville et al.\ 2007) as they sample up to millions of galaxies down to AB magnitudes fainter than $26$, rendering a spectroscopic follow-up of {\em all} sources impossible. 
Such surveys, however, usually contain excellent (X-ray to radio) multi-wavelength coverage which allows selecting radiatively efficient AGN (i.e., HERG-dominated samples) using  X-ray and IR data (e.g. Brusa et al. 2007;           Donley et al. 2013), and radio-loud AGN (i.e.\ LERG dominated samples) as outliers in the IR-radio correlation (e.g. Condon 1992; Bonzini et al. 2013; Delvecchio et al., in prep.). 
Furthermore, fitting the densely covered optical to IR spectral energy distribution (SED) of sources using sophisticated spectral libraries allows to approach the separation via i) rest-frame colors which have been shown to correlate with the position of a galaxy in the spectroscopic diagnostics diagram (e.g. Baldwin et al.\ 1981; \smo \ et al. 2006, 2008), and/or ii) SED fitting using both galaxy and AGN templates and, thereby, identifying sources with substantial AGN contribution, as well as determining the fractional contribution of the AGN to the galaxies' bolometric luminosity (Delvecchio et al.\ 2014, in prep). 

In summary, in the literature  radio sources drawn from large-area, but shallow radio surveys (e.g., NVSS, FIRST) are usually combined with large, and complete spectroscopic surveys (e.g., SDSS, 2dF, GAMA, see e.g., \smo \ 2009; Sadler et al.\ 2003; Prescott et al.\ 2016) and, thus, separated into star forming and AGN galaxies based on optical spectroscopic diagnostic tools (e.g., Best et al.\ 2005; Sadler et al.\ 2002; Donoso et al.\ 2009; \smo \ 2009), reaching out to $z\sim1$. On the other hand, radio sources drawn from intermediate- and small-area, but deeper radio surveys (e.g., COSMOS, E-CDFS; Schinnerer et al.\ 2007; \smo \ et al., in prep.; Miller et al.\ 2008, 2013) are usually separated using multi-wavelength data as proxies for SF- and AGN-related processes (e.g., \smo \ et al. 2008, Bonzini et al.\ 2013, Padovani et al.\ 2015, Baran et al., in prep; Delvecchio et al., in prep), and currently reach out to $z\sim5$. 

\begin{figure}
\begin{center}
\includegraphics[bb=  54 460 486 652, scale=0.8]{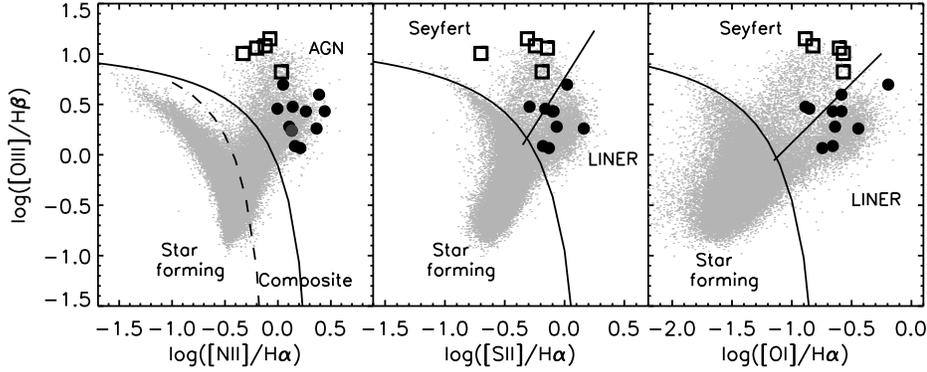}
\caption{ Optical spectroscopic diagnostic diagrams (see Kauffmann et al.
2003b; Kewley et al. 2006) that separate emission-line galaxies into star forming,
composite galaxies, and Seyfert and LINER AGN.
Small gray dots represent galaxies from the SDSS DR4 "main" spectroscopic
sample. Large open squares (filled dots) denote $z < 0.1$ 3CRR (Laing et al.\ 1983) radio galaxies 
independently classified (based on their core X-ray emission) as systems with
radiatively efficient (inefficient) black hole accretion (Evans et al. 2006). Adapted from \smo \ (2009).}
\label{fig:bpt}
\end{center}
\end{figure}

\section{Properties of radio AGN}
\label{sec:props}

\subsection{Low redshift}

To investigate the principal differences in physical properties of LERAGN and HERAGN \smo \ (2009)
 used  a unified catalog
of radio objects detected by NVSS, FIRST, WENSS, GB6, and
SDSS (Kimball \& Ivezi\'{c} 2008). They augmented this catalog
with derivations of emission-line fluxes, 4000~\AA \ breaks, stellar
masses, and stellar velocity dispersions, drawn from the SDSS DR4
"main" spectroscopic sample (see \smo \ 2009 and references therein), and further limited it to  unique objects that have been detected
by both the FIRST and NVSS surveys at 20 cm and with redshifts $0.04 < z < 0.1$. Separating the objects into Seyfert, LINER, and absorption line systems based on their spectroscopic emission line properties (see previous section, and \f{fig:bpt} ), they found a clear dichotomy  between the properties of low-excitation (absorption line AGN and LINERs) and high-excitation (Seyferts) radio AGN. This is summarized in \f{fig:lowzprops} \  where the distributions of host galaxy color, 4000~\AA \ break strength, stellar mass, as well as proxies for the SMBH mass and accretion rate are shown for the various AGN populations. In summary, they find that hosts
of LERAGN have the highest stellar masses, reddest optical colors, and highest mass black holes but accrete radiatively 
inefficiently, and at low accretion rates. On the other hand, the high-excitation radio AGN have lower stellar masses, bluer
optical colors (consistent with the "green valley"), and lower mass black holes that accrete radiatively efficiently, and at
high ($\sim$~Eddington) accretion rates. Note that Best \& Heckman (2012) find consistent  results using a similar sample of FIRST-NVSS-SDSS galaxies, separated into HERAGN and LERAGN. The dichotomy between the two AGN populations can be summarized in the galaxy color vs.\ stellar mass plane, as shown in \f{fig:cmd} . 
 In the local universe LERAGN occupy the red sequence of galaxies, while HERAGN populate the green valley. 
 This is also consistent with a systematic difference found for average gas masses within the host galaxies of the two AGN populations. 
 \smo \ \& Riechers (2011) performed CO(1$\rightarrow$0) observations with the CARMA interferometer towards 11 $z<0.1$ radio AGN drawn from the 3CRR catalog (Laing et al.\ 1983; Ewans et al.\ 2006), and separated into LERAGN and HERAGN based on their spectroscopic emission line properties (Buttiglione et al.\ 2009).  Combining these results with literature values they derived  molecular gas masses (or upper limits) for a complete sample of 21 $z < 0.1$ radio AGN, finding, on average, a factor of $\sim7$ higher gas masses within the host galaxies of HERAGN, compared to LERAGN.
 
In the context of the blue-to-red galaxy evolution scenario the above described results  imply that  HERAGN occur in an earlier stage of galaxy evolution relative to LERAGN. Furthermore, HERAGN are hosted by galaxies with intrinsically higher SF rates (occupying the green valley) than LERAGN (occupying the red sequence). Hence, as, in general, radio emission in the observed (cm) band predominantly arises from synchrotron emission in galaxies caused either by SF- (supernovae explosions) or AGN-related processes (see Condon 1992 for a review), this naturally leads to the question about  the real source of the observed radio emission in HERAGN and LERAGN. Note that the selection of these two AGN types is based on their optical emission line properties indicating either the existence or lack of a radiatively efficient AGN powerful enough to excite the given emission lines. However, even if a powerful AGN is identified this way its observed emission in the radio band may not necessarily fully arise from the AGN. If the host galaxy is forming stars at a substantial rate this will, per-definition, contribute to the observed radio emission of the source, rendering the radio emission a  mix of SF- and AGN-induced radiation. Mori\'{c} et al.\ (2010) have inferred the statistical fractional contribution of SF- and AGN-related processes contributing to the observed radio emission in HERAGN and LERAGN using a sample of SDSS-NVSS-IRAS galaxies at $0.04<z\lesssim0.2$, separated into Seyfert, LINERs, and absorption line galaxies based on their spectroscopic line properties. They derived this quantity by assessing the average offset from the  IR-radio correlation (well-known to firmly hold for SF galaxies; e.g., Condon 1992) for a specific AGN/galaxy population (see Mori\'{c} et al.\ 2010 for details). They find that, statistically, about 60\% of the radio emission in HERAGN arises from SF-, rather than AGN-related processes, while for LERAGN only about 10\% of the radio emission accounts for SF-related processes (see also G{\"u}rkan et al.\ 2015, Hardcastle et al.\ 2013). Note that for the more powerful radiatively efficient AGN -- quasars --  the source of radio emission in radio-quiet quasars is still debated (see Kimball et al 2011; Condon et al. 2013, White et al. 2015). 
 
Lastly, the local radio luminosity functions derived separately for HERAGN and LERAGN have been presented in e.g.\ Filho et al.\ (2006), Best \& Heckman (2012), Gendre et al. (2013), Pracy et al.\ (2015; in prep). While it is clear that HERAGN dominate the volume densities at high radio luminosities (\lum~$>10^{26}$~\wh ; Best \& Heckman 2012), the slope of the low-luminosity end of the HERAGN radio luminosity function is still unclear (e.g., Best \& Heckman 2012 find a significantly flatter slope than Filho et al.\ 2006; see also Pracy et al., in prep; Padovani et al.\ 2015). This is most likely due to the difficulty of disentangling the real contribution of AGN-related radio emission in HERAGN. 

In summary, in the local universe fundamental physical differences have been found between LERAGN and HERAGN. LERAGN occupy the red sequence of galaxies, they are hosted by the highest stellar mass galaxies, and contain the most massive SMBHs accreting at low, sub-Eddington rates in a radiatively inefficient manner (consistent with accretion dominated advection flows -- ADAFs), and they are predominant at lower radio luminosities (\lum~$<10^{26}$~\wh ). On the other hand, HERAGN ccupy the green valley of galaxies, they are hosted by less massive galaxies, and contain less massive SMBHs, but accrete at high, Eddington rates in an radiatively efficient manner (consistent with a thin-disk geometry), and they are predominant at higher radio luminosities (\lum~$>10^{26}$~\wh ).

\begin{figure}
\begin{center}
\includegraphics[bb=  30 628 612 752, scale=0.9]{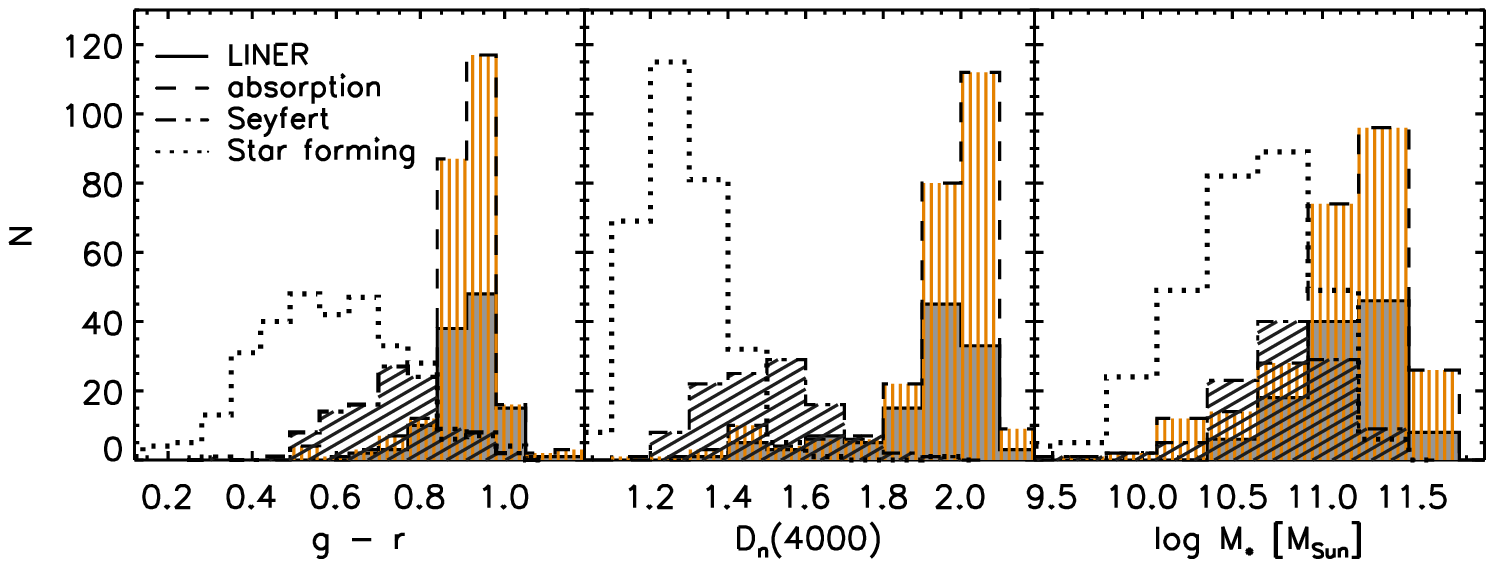}
\includegraphics[bb= 40 400 612 622, scale=0.9]{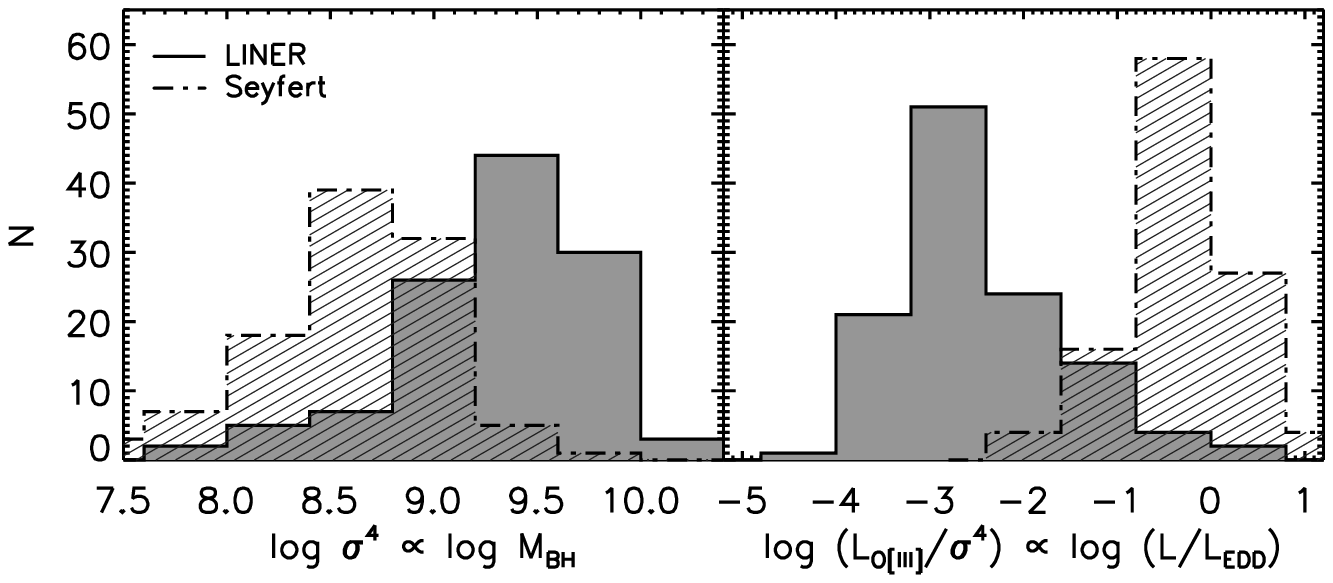}
\caption{From left to right the top panels show the distribution of the i) observed $g-r$ color, ii) 4000~\AA \ break strength, and iii) stellar mass, while the bottom panels show the distribution of the iv) velocity dispersion, $\sigma^4$,
proportional to black hole mass (Tremaine et al. 2002), and v) $\log ({L_\mathrm{[OIII]} / \sigma^4} )$ proportional to black hole accretion rate (in Eddington units;
Heckman et al. 2004) for various galaxy populations  drawn from the FIRST-NVSS-SDSS sample (star forming galaxies: dotted lines and empty histograms; Seyferts: Dash-dotted lines and diagonally hatched histograms; absorption line galaxies: dashed lines and vertically hatched histograms; LINERs: full lines and filled histograms), and also  indicated in the panels. Adapted from \smo \  (2009).}
\label{fig:lowzprops}
\end{center}
\end{figure}

\begin{figure}
 \centerline{\begin{minipage}[l]{6.9cm}
\includegraphics[bb=  20 0  588 497, scale=0.35]{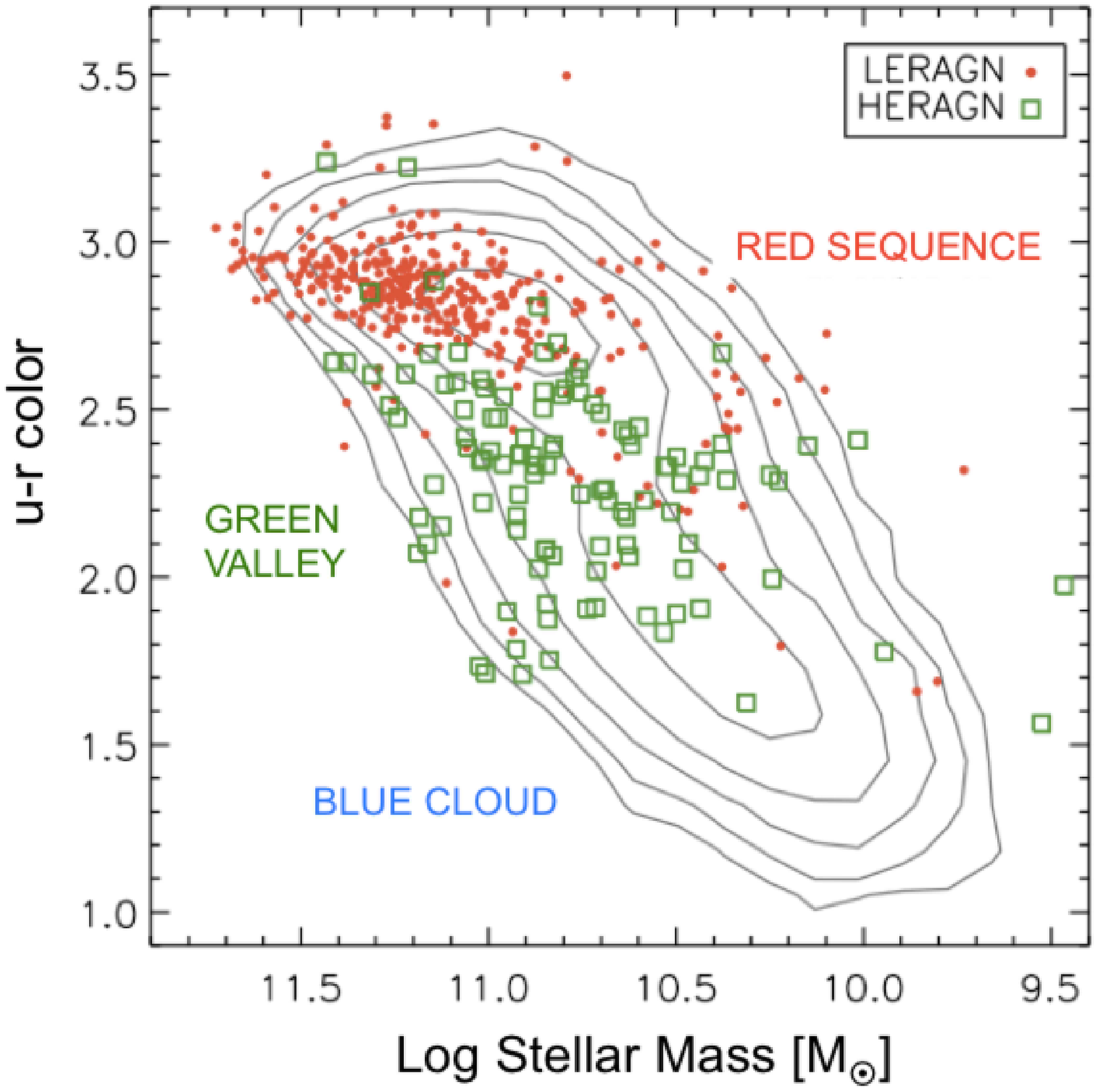}
\caption{Observed $u-r$ color as a function of stellar mass for FIRST-NVSS-SDSS galaxies separated into LERAGN (LINER+absorption line galaxies), and HERAGN (Seyferts) using spectroscopic emission line diagnostics (Baldwin et al.\ 1981; Kauffmann
et al. 2003; Kewley et al. 2001, 2006). Adapted from \smo \ (2009).}
\label{fig:cmd}
      \end{minipage} \ \hfill \ 
  \begin{minipage}[l]{7.6cm}
\includegraphics[bb= 0 0 371 357, scale=0.55]{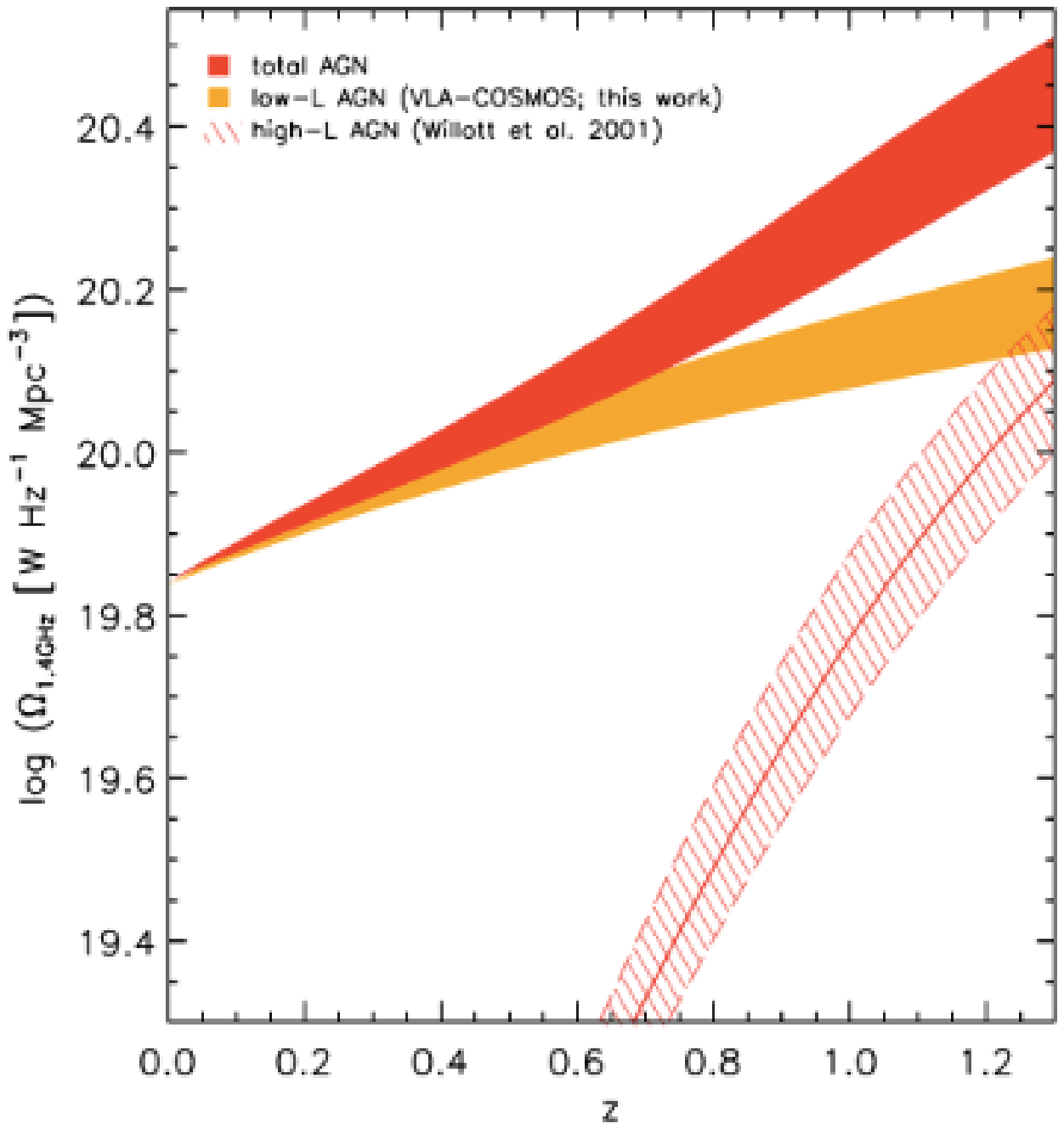}
\caption{Evolution of the comoving 20 cm integrated luminosity density for
VLA-COSMOS AGN (orange curve) galaxies since $z = 1.3$. Shown is also the
evolution of  high-luminosity radio AGN, adopted from Willott et al. (2001,
hatched region). The evolution for the total, co-added AGN population is shown as the red-shaded curve. Adapted from \smo \ et al.\  (2009).}
\label{fig:evolv}
     \end{minipage}} 
\end{figure}

\subsection{High redshift}

Investigating physical properties of HERAGN- and LERAGN-dominated samples in the high redshift universe ($z>1$) is the focus of current studies. Detailed studies of this kind were enabled only recently with the advent of deep radio continuum surveys with excellent multi-wavelength coverage allowing robust separations of radio AGN into various galaxy populations, but also covering enough area to allow construction of statistically significant samples of distinct types of radio AGN, divided in bins of various properties such as redshift, stellar mass, (specific) SF rate etc. 

Bonzini et al.\ (2015) studied  SF properties of $\sim800$ radio sources drawn from the E-CDFS survey at 1.4GHz (0.3 deg.$^2$, flux density limit of 32.5 $\mu$Jy at the field center; Miller et al.\ 2008, 2013), separated into SF, radio-loud and radio-quiet AGN (Bonzini et al.\ 2013). As discussed in Bonzini et al.\ (2013) and Padovani et al.\ (2015) their sample of radio-quiet AGN is expected to be dominated by HERAGN, while their radio-loud AGN sample is expected to predominantly contain LERAGN. Bonzini et al.\ (2015) find that  the radio luminosity of their radio-quiet AGN ($z\sim1.5 - 2$)  is consistent with the galaxies' SF rates inferred using their FIR luminosites. 
The first results based on the VLA-COSMOS 3GHz Large Project (2.6 sq.degs.; $rms=2.3~\mu$Jy/beam; $\sim11,000$ source components, \smo \ et al., in prep.) are consistent with this. These studies (Delhaize et al., in prep; Delveccio et al., in prep.) rely on a sample of $\sim 1,500$ radiatively efficient AGN ($z<5$; drawn from the survey and cross-correlated with the COSMOS multi-wavelength data set; Baran et al., in prep), which were identified using their X-ray emission ($L_\mathrm{X}>10^{42}$~erg/s; Civano et al., subm.), IR colors (using the criteria defined by Donley et al.\ 2008), and SED fitted properties (using both galaxy and AGN templates; see Delvecchio et al. 2014 for details). Performing a similar analysis as that presented in Mori\'{c} et al. (2010; see previous Section), Delhaize et al. (in prep.) infer that a substantial fraction of the radio emission in HERAGN ($z<5$) may arise from SF-related, rather than AGN-related processes. This is consistent with the SED properties of this sample as inferred by Delvecchio et al.\ (in prep), who however also find that about 35\% of HERGs show a $ >3\sigma$ radio-excess compared to the IR-based SF rate. 
Delvecchio et al. (in prep.) further investigate the physical properties of radiatively efficient AGN (HERAGN-dominated), as well as radiatively inefficient AGN (LERAGN-dominated) defined as those galaxies that have not been selected using the X-ray, IR, and SED criteria (as defined above), but exhibit an offset in radio luminosity relative to that expected based on the SF rate inferred in the host galaxies via their IR emission. 
They find that, relative to those of HERAGN, the distributions of stellar mass 
and specific SF rates of LERAGN are, respectively, skewed towards higher and lower, values out to $z\sim1$. Beyond this redshift the distributions appear to become similar, suggesting a change in host galaxy properties at these higher redshifts. 

In summary, evidence has been found supporting that the radio emission in HERAGN dominated samples (X-ray-, IR-, and SED-selected AGN) substantially arises from SF processes  at high ($z>1$) redshift, consistent with the findings for low-redshift HERAGN (see previous Section). Studies of the physical properties of the various radio AGN populations, and their potential dichotomy across cosmic time are underway. 

\section{Evolution of radio AGN}
\label{sec:evolv}

Past studies have shown that radio luminous AGN evolve via a "down-sizing" effect, i.e. low-luminosity sources evolve less strongly than high-luminosity sources (e.g., Willott et al.\ 2001; Rigby et al.\ 2015). 
Studies of high-luminosity radio AGN (\lum~$\gtrsim 2 \times 10^{26}$~\wh ; Dunlop \& Peacock 1990; Willott et al. 2001) have found a strong positive density evolution at $z\lesssim2$, beyond which their comoving volume density declines. A substantially slower evolution of this population, with a  lower redshift ($z\sim 1-1.5$) comoving volume density turnover, has been found for lower-luminosity radio AGN (\lum~$ > 2 \times 10^{25}$~\wh ; Waddington et al. 2001). Studies of even lower luminosity radio AGN (\lum~$\lesssim10^{25}$~\wh ) find a mild evolution out to $z\sim1$ (e.g., \smo  \ et al.\ 2009). Combining these results in context of the cosmic evolution of radio AGN samples dominated by HERAGN and LERAGN suggests a much stronger cosmic evolution of the first compared to the latter, as illustrated in \f{fig:evolv} . Direct measurements of the cosmic evolution of HERAGN and LERAGN, separated based on spectroscopic emission line properties out to $z\sim1$ are in qualitative agreement with these results (Sadler et al.\ 2007; Donoso et al\ 2009; Best et al.\ 2014). 
Expanding this to higher redshifts Padovani et al.\ (2015) have constrained the cosmic evolution of their radio-loud (LERAGN-dominated) and radio-quiet (HERAGN-dominated) AGN identified in the E-CDFS survey out to $z\sim4$ based on a sample of $\sim300$ radio AGN. While they find a strong evolution (similar to that for SF galaxies) of the first sample throughout the redshift space probed, they find that the number density of the second peaks at $z\sim0.5$, and declines at higher redshifts.

\section{Summary and outlook}
\label{sec:summ}

A dichotomy reflecting fundamental physical differences between low- and high- excitation radio AGN has been shown to exist in the local Universe. This is summarized in \f{fig:props} , and relates to stellar, gas and SMBH mass, accretion rate and mode, cosmic evolution, as well as the source of radio emission. Evidence has been found in support of the observed radio emission in HERAGN substantially arising from star-formation, rather than AGN-related processes, both in the local and high-redshift Universe. 
While numerous studies have focused, and converged on many properties of low-redshift HE- and LERAGN, the physical properties of such sources at higher redshifts are the focus of current studies. This is enabled by current radio continuum surveys probing to-date the faintest levels, in conjunction with excellent multi-wavelength coverage. The results of these will set the path for planned surveys, such as those with the SKA and its precursors.

\begin{figure*}
\begin{center}
\includegraphics[bb= 15 0  562 517, scale=0.8]{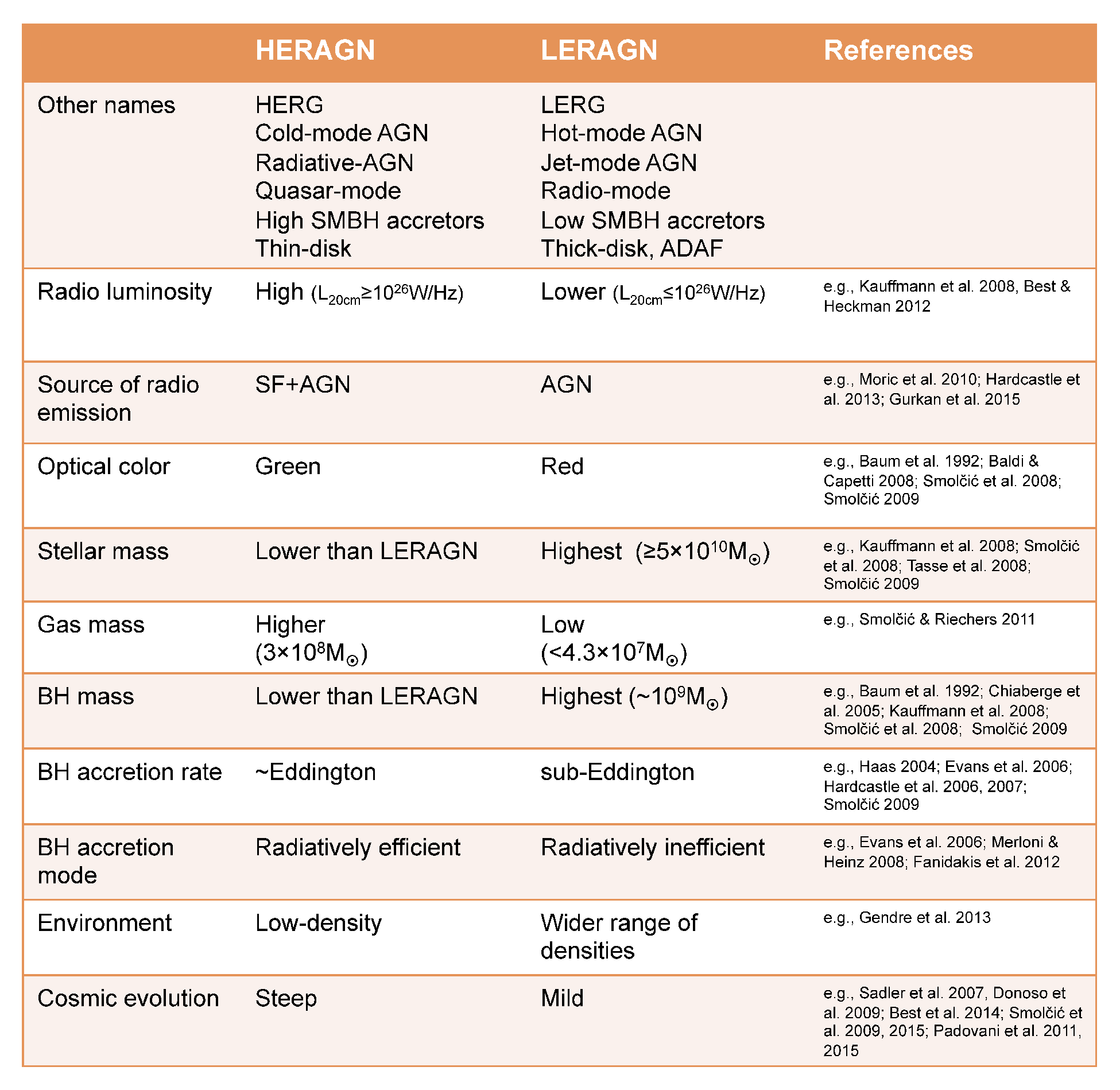}
\caption{Summary of properties of high-excitation and low-excitation radio AGN (HERAGN and LERAGN, respectively). }
\label{fig:props}
\end{center}
\end{figure*}

\end{document}